# Properties of Random Complex Chemical Reaction Networks and Their Relevance to Biological Toy Models

Erwan Bigan[1,2], Jean-Marc Steyaert[1] and Stéphane Douady[2]

[1] Laboratoire d'Informatique (LIX), École Polytechnique, 91128 Palaiseau Cedex, France

[2] Laboratoire Matière & Systèmes Complexes, UMR7057 CNRS, Université Paris Diderot, 75205 Paris Cedex 13, France

erwan.bigan@m4x.org

**Abstract** *We investigate the properties of large random conservative chemical reaction networks composed of elementary reactions endowed with either mass-action or saturating kinetics, assigning kinetic parameters in a thermodynamically-consistent manner. We find that such complex networks exhibit qualitatively similar behavior when fed with external nutrient flux. The nutrient is preferentially transformed into one specific chemical that is an intrinsic property of the network. We propose a self-consistent proto-cell toy model in which the preferentially synthesized chemical is a precursor for the cell membrane, and show that such proto-cells can exhibit sustainable homeostatic growth when fed with any nutrient diffusing through the membrane, provided that nutrient is metabolized at a sufficient rate.*

## 1 Introduction

Toy models help understand minimal requirements of life. Significant research has already been devoted to this field using random chemical reaction networks [1, 2, 3]. We build upon this previous work, extending this approach to random mass conservative chemical reaction networks with arbitrary stoichiometry.

Considering an arbitrary set of chemicals with randomly assigned standard Gibbs free energies of formation, random conservative networks of elementary reactions are generated, and kinetic parameters are randomly assigned to direct reactions, while kinetic parameters for reverse reactions are derived in a thermodynamically consistent manner. In addition to conventional mass-action kinetics, we also consider saturating kinetics. This is because mass-action kinetics only hold for diluted media, whereas biological systems are typically dense and macromolecular crowding is known to reduce effective reaction rates in the diffusion-limited regime [4].

This paper is organized as follows: in section 2, we describe our random reaction network model; in section 3, we characterize generated reaction networks in terms of equilibrium concentrations versus total mass density (section 3.1) and in terms of behavior under external nutrient flux (section 3.2); in section 4, we describe a proto-cell model enclosing such random reaction networks within a membrane (section 4.1) and explore the range of parameters within which cellular growth can be sustained (section 4.2); section 5 is devoted to discussion and conclusion.

## 2 Random Conservative Reaction Network Model

N different chemicals $\{A_i\}_{i=0, ..., N-1}$ are present in the system. Reactions are decomposed in monomolecular or bimolecular elementary reactions:

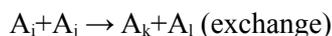 $A_i + A_j \rightarrow A_k + A_l$ (exchange)

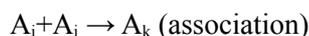 $A_i + A_j \rightarrow A_k$ (association)

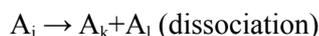 $A_i \rightarrow A_k + A_l$ (dissociation)

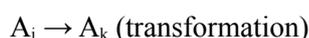 $A_i \rightarrow A_k$ (transformation)

It can be easily shown that any reactions with higher-order stoichiometry can be decomposed in such elementary reactions (at the expense on increasing the total number of reactions, and of increasing the total number of chemicals – typically intermediate composite compounds). Reactions with i=j or i=k are allowed, the only constraint is that the same expression should not be found on both sides.

All results presented in this paper have been obtained (i) excluding exchange reactions from the set, considering that they can be represented as successive combinations of association and dissociation reactions; and (ii) using N=10 as total number of chemicals (similar qualitative behavior has been observed with N=20, at the expense of longer computational time).

## 2.1 Topology

We use a previously published standard algorithm to check whether a network is conservative or not [5]. Running this algorithm also delivers the generating vectors of the convex cone of admissible mass vectors (a vector m with components $\{m_i\}_{i=0, ..., N-1}$ is an admissible mass vector if and only if $m^T S=0$ where $^T$ denotes the transpose operation and S is the N x R stoichiometry matrix, with R being the total number of reactions). To generate a random chemical reaction network, we proceed as follows: we choose a first random reaction, and then successively add new randomly chosen reactions while checking at each step that the network is conservative.

We observe that once the number of reactions exceeds N by a handful, the number of generating vectors of the convex cone falls to one, i.e. up to a multiplying factor there is only one single admissible mass vector for the given reaction network. At significantly larger number of reactions, we also observe that conservative networks reach a maximum size above which none of the remaining elementary reactions is orthogonal to the single mass vector, and that this maximum size is variable (ranging from 17 to 96 with an average of 42 direct reactions for a sample of 100 networks). Figure 1 shows an example of such a maximum-sized random conservative chemical reaction network that will be used as reference example in the remainder of this paper.

## 2.2 Kinetics

Standard Gibbs free energies of formation $\{G_i\}_{i=0, ..., N-1}$ are assigned to chemicals $\{A_i\}_{i=0, ..., N-1}$. For a given possible reaction of the constructed network, the forward versus reverse directions are determined by comparing the sum of standard Gibbs free energies of formation for reactants versus products.

Mass-action kinetics are characterized by a kinetic coefficient $k_r$ such that the reaction rate $f_r$ is given by $f_r=k_r[A_i][A_j]$ (resp. $f_r=k_r[A_i]$) for a bimolecular (resp. monomolecular) reaction. $k_r^\rightarrow$ are first assigned to direct reactions, and $k_r^\leftarrow$ for reverse reactions are then derived as given in the expressions below, where $\Delta G$ denotes the change in Gibbs free energy for the considered reaction, R the ideal gas constant, and c° the standard concentration:

- Monomolecular forward ($A_i \rightarrow ...$): $k_r^\rightarrow = k_{avg\_monomolecular} 10^{random(-s/2, s/2)}$
    - Monomolecular reverse ($A_i \leftarrow A_k$): $k_r^\leftarrow = k_r^\rightarrow \exp(-|\Delta G|/RT)$
    - Bimolecular reverse ($A_i \leftarrow A_k+A_l$): $k_r^\leftarrow = (k_r^\rightarrow/c°)\exp(-|\Delta G|/RT)$
- Bimolecular forward ($A_i+A_j \rightarrow ...$): $k_r^\rightarrow = k_{avg\_bimolecular} 10^{random(-s/2, s/2)}$
    - Monomolecular reverse ($A_i+A_j \leftarrow A_k$): $k_r^\leftarrow = c° k_r^\rightarrow \exp(-|\Delta G|/RT)$
    - Bimolecular reverse ($A_i+A_j \leftarrow A_k+A_l$): $k_r^\leftarrow = k_r^\rightarrow \exp(-|\Delta G|/RT)$

with random(-s/2, s/2) designating a random real number between -s/2 and s/2 where s is the spread of kinetic parameters for direct reactions, counted in orders of magnitude.

Saturating kinetics are simply derived from mass-action kinetics by introducing a saturation concentration $K_r$ for each reaction (chosen independently for forward and reverse reactions), and by modifying mass-action kinetics the following way:

- Monomolecular: $f_r=k_r\{[A_i]/(1+[A_i]/K_r)\}$

- Bimolecular: $f_r = k_r \{[A_i]/(1+[A_i]/K_r)\} \{[A_j]/(1+[A_j]/K_r)\}$

where $K_r = K_{avg} 10^{random(-p/2, p/2)}$, with p the spread of saturation concentration counted in orders of magnitude. Saturating kinetics tend towards mass-action kinetics at low concentrations, as should be expected.

Chosen numerical values were $k_{avg\_monomolecular} = 10^2\,s^{-1}$, $k_{avg\_bimolecular} = 10^4\,M^{-1}.s^{-1}$ (typical $k_{cat}$ and $k_{cat}/K_m$ for enzymatic reactions), and $K_{avg} = 10^{-2}\,M$. Standard Gibbs free energies of formation were assigned as $G_i/RT = random(0, 15)$ so that the maximum change in Gibbs free energy approaches that of ATP hydrolysis. Similar qualitative behavior were obtained irrespective of (s, p) values over several orders of magnitude of spreads. Results presented in the following were obtained with s=p=0.

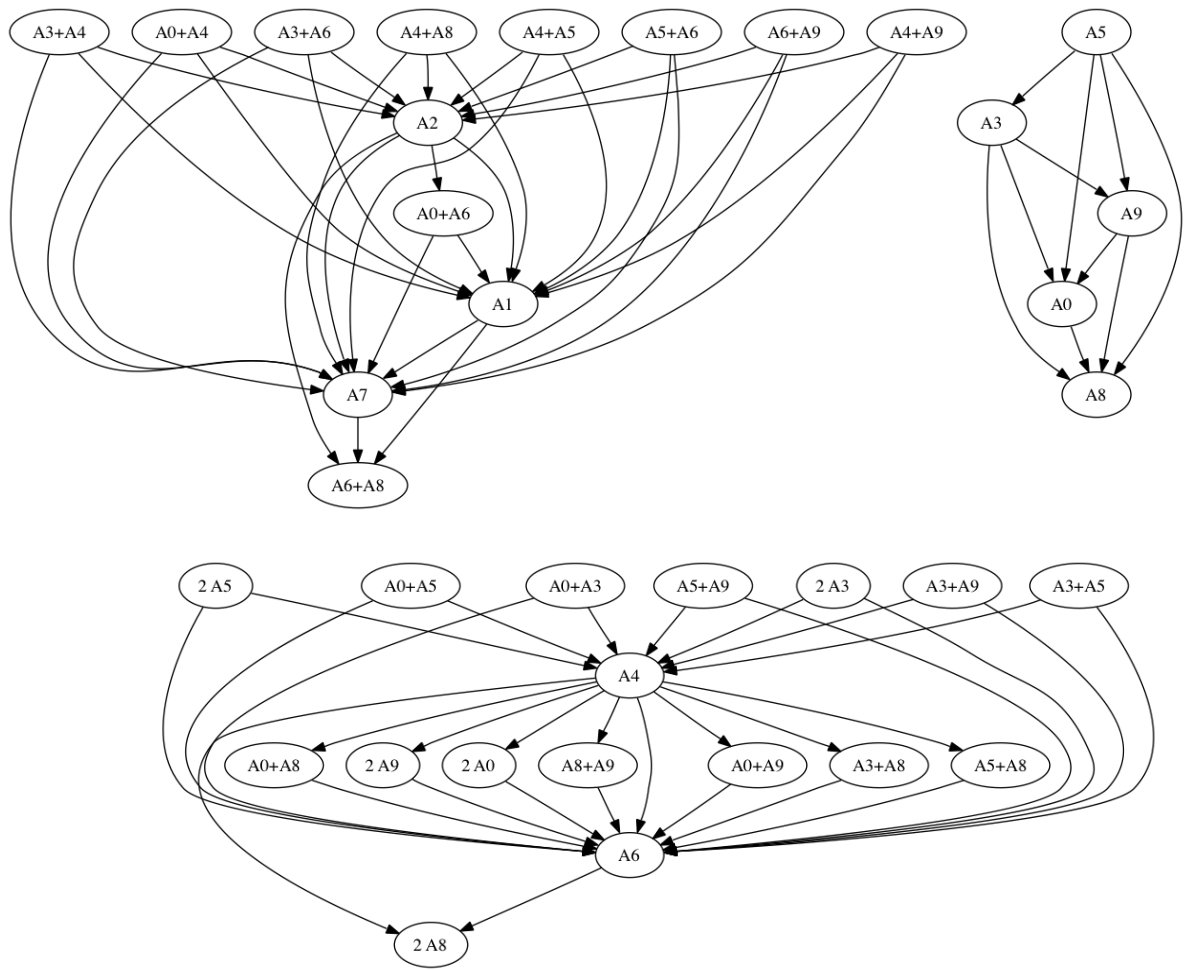

**Figure 1.** Example of a maximum-sized random conservative chemical reaction network, with N=10. Maximum size was reached for 74 direct reactions represented as arrows (148 reactions when counting both direct and reverse). Single mass vector components for this network are $\{m_i\}_{i=0,...,9} = \{1, 3, 3, 1, 2, 1, 2, 3, 1, 1\}$.

# 3 Characterization of Random Reaction Networks

## 3.1 Equilibrium Behavior

The dynamics of the closed system are governed by the following equation:

$$dA/dt = Sf$$

where A is the N-vector of concentrations for the different chemicals, S is the NxR stoichiometry matrix, and f is the R-vector of reaction flux for the different reactions (components $f_r$ of vector f being a function of A through the kinetics given in the previous section).

Equilibrium concentrations were determined for different values of the system density D ($D=\sum_i m_i[A_i]$) by computing the concentration trajectories for sets of initial conditions with increasing concentrations. Unicity of the equilibrium for a given total density was checked by comparing computed equilibria under different initial conditions of same total density (equal distribution among chemicals vs. allocation to a single chemical, this process being repeated for each of the N chemicals).

Detailed balance at equilibrium is guaranteed from thermodynamic theory with mass-action kinetics [6], but not with saturating kinetics. Indeed, we observe that detailed balance is verified at any density with mass-action kinetics, but not for saturating kinetics: when the equilibrium deviates significantly from that with mass-action kinetics, some reactions are detailed-balanced while others are not.

Figure 2 gives equilibrium concentrations as a function of system density. At sufficiently low system density for which no reaction saturates, identical equilibrium concentrations are obtained with mass-action or saturating kinetics. At higher density, the behavior differs significantly with all the extra mass in the system going to a single chemical in the case of saturating kinetics.

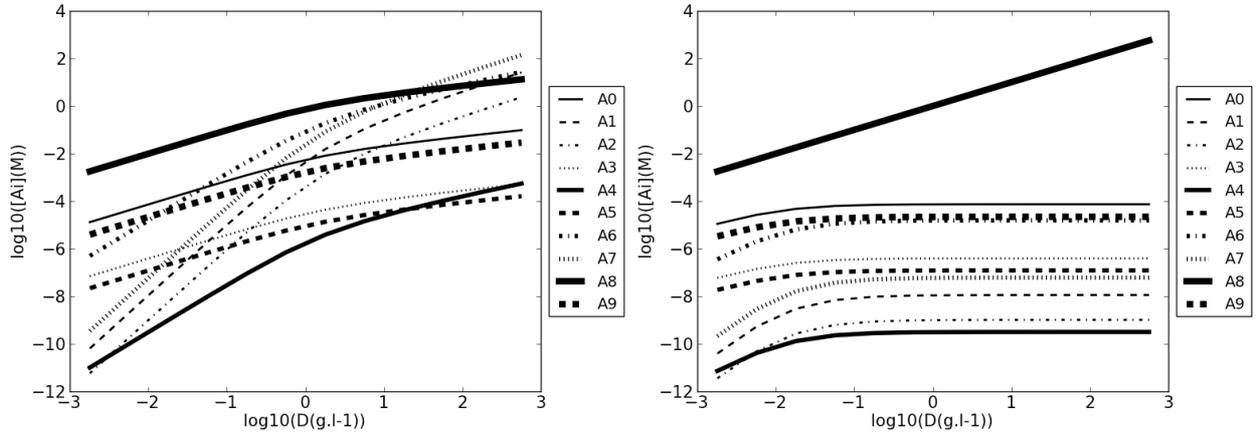

**Figure 2.** Equilibrium concentrations as a function of system density for an example randomly generated maximum-sized conservative system, with mass-action kinetics (left) and with saturating kinetics (right).

## 3.2 Behavior under External Nutrient Flux

We now consider system behavior when submitted to an external nutrient flux, $f_{nu}$, of any one of the N different species. The dynamics of the system are governed by the following equation:

$$dA/dt = Sf + f_{nu}$$

where $f_{nu}$ is the N-input flux vector having all components null except for the one injected chemical.

The rate of mass increase is conserved because the system is conservative: $m^T S = 0 \Rightarrow m^T dA/dt = m^T f_{nu}$, or equivalently: $\sum_{j=0,\ldots,N-1} m_j d[A_j]/dt = m_{nu} f_{nu}$.

We first consider saturating kinetics. Figure 3 shows a typical trajectory with $A_5$ arbitrarily chosen as nutrient. While dynamics on a short time scale (left) exhibit complex behavior, on a longer time scale (right) all chemicals except $A_8$ asymptotically reach constant non-zero concentrations, while all the injected mass is transformed into $A_8$, with $[A_8]$ diverging linearly asymptotically. This asymptotic behavior is independent of initial conditions. Using different nutrients leads to the same preferential transformation into $A_8$. In essence, the system acts as a directed transformation machine.

To further investigate this behavior, we have computed trajectories over a wide range of input flux, and up to a sufficiently large fixed given time. Figure 4 shows on log scales the values of densities $m_i[A_i]$ (top) and of their derivatives $m_i d[A_i]/dt$ (bottom) after the fixed given integration time.

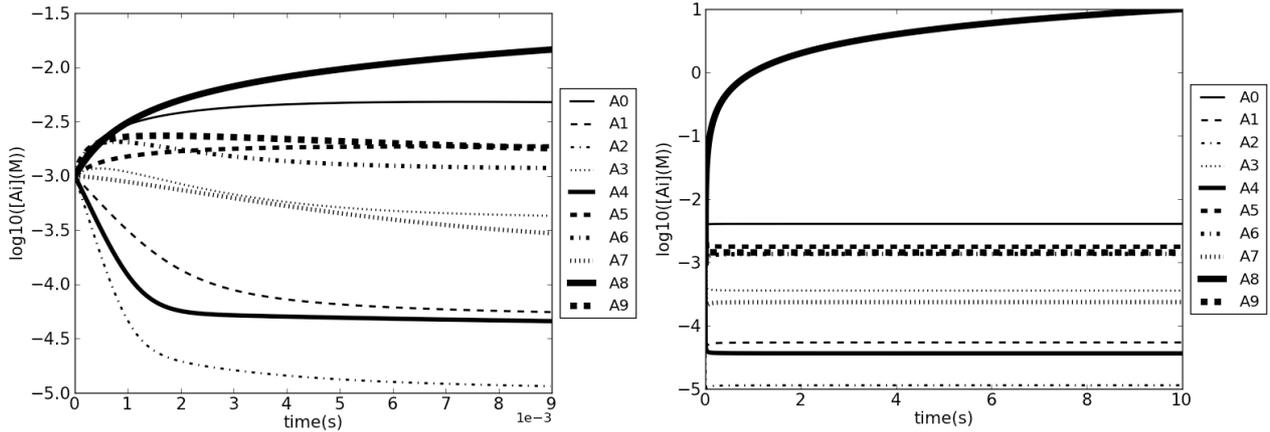

**Figure 3.** Concentrations vs. time on short (left) and long (right) time scales with a constant input flux $f_{nu}=1$ $M.s^{-1}$ of nutrient $A_5$, for saturating kinetics. Initial conditions are 1 mM for all species.

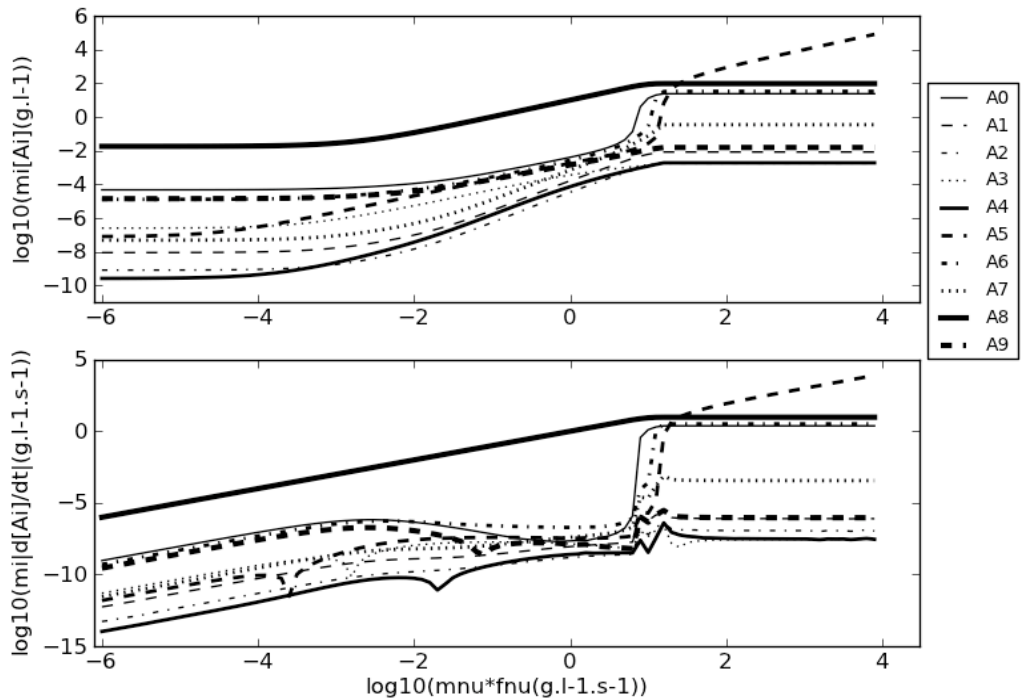

**Figure 4.** Densities $m_i[A_i]$ (top) and their derivatives $m_i d[A_i]/dt$ (bottom) after 10 s integration time, vs. nutrient density flux, for saturating kinetics. Nutrient is $A_5$. Initial conditions are 1 mM for all species.

Figure 4 (top) shows that at low $f_{nu}$ the injected mass is too low to significantly change equilibrium concentrations, even after the 10 s integration time. Above some threshold $f_{nu}$ (that depends on the system density, and thus on the integration time), $A_8$ becomes significantly more abundant. Consistently, Figure 4 (bottom) shows that at low $f_{nu}$ the conserved rate of mass increase $m_{nu}f_{nu}=\sum_{j=0,...,N-1} m_j d[A_j]/dt$ is distributed among chemicals, while above the threshold $f_{nu}$ it nearly all goes to $A_8$. In this regime, asymptotical trajectories are typically as shown on Figure 3: $A_8$ diverges linearly and all other chemicals including the injected nutrient reach constant non-zero values.

The same directed transformation behavior as shown on Figures 3 and 4 is observed independently of the chosen nutrient, the one particular preferentially synthesized chemical ($A_8$ in the present example) being an intrinsic property of the system.

At even higher $f_{nu}$, a different regime is reached with the transformation capacity of the system being saturated, and all the extra added mass remaining intact. In this saturated regime, most reactions are saturated, and all reaction rates are locked independently of the $f_{nu}$ value.

When using mass-action instead of saturating kinetics, the same threshold behavior is observed, but all concentrations increase with increasing $f_{nu}$, and as expected no saturation regime is reached.

We have generated and characterized tens of random maximum-sized conservative systems, and observed that they all qualitatively behave the same way. The preferentially synthesized species depends on both network topology and kinetics, and is statistically a heavier lower-energy chemical.

## 4  Proto-Cell Model

We have found that most large conservative reaction networks with saturating kinetics effectively function as directed transformation machines, preferentially converting any nutrient into one particular chemical over a wide range of input fluxes. This particular chemical is an intrinsic property of the reaction network and is also the most abundant chemical in the system.

Most abundant chemicals in actual biological systems are typically structural molecules, first of which membrane molecules or their precursors. It is thus tempting to assign such a role to this most abundant and preferentially synthesized species in our model, which we will denote $A_{me}$.

### 4.1  Membrane Model

In the following, we assume that $A_{me}$ meets the membrane requirements: (i) ability to self-assemble in a continuous membrane and (ii) permeability of the self-assembled membrane to nutrients $A_{nu}$.

Regarding (i), we further assume $A_{me}$ is incorporated into the growing membrane at a molar rate per unit area: $\mathcal{F}_{me}=\mathcal{K}_{me}[A_{me}]/(1+[A_{me}]/K_{me})$ with $\mathcal{K}_{me}$ being a kinetic rate per unit area and $K_{me}$ a saturation concentration. This results from the unidirectional reaction $A_{me} \rightarrow A_{me\_structured}$ (once assembled, membrane constituents no longer react with other chemicals). Molar rate per unit volume $f_{me}$ (resp. kinetic rate $k_{me}$ per unit volume) are simply derived multiplying $\mathcal{F}_{me}$ (resp. $\mathcal{K}_{me}$) by the (*Area*/*Volume*) ratio: $f_{me}=k_{me}[A_{me}]/(1+[A_{me}]/K_{me})$.

As new $A_{me\_structured}$ get incorporated in the membrane, the membrane *Area* increases at a relative rate equal to $\mathcal{F}_{me}$ divided by the number of molecules per unit membrane area Nmea, that is an intrinsic property of the self-assembling molecules: $(1/Area)dArea/dt=\mathcal{F}_{me}/$Nmea.

By definition, cellular growth rate $\mu$ is the relative rate of volume increase, and is directly related to the relative rate of area increase by a multiplying factor (3/2 for a sphere, 1 for a filament). Assuming a filament shape for simplicity gives: $\mu=\mathcal{F}_{me}/$Nmea$=f_{me}/$[membrane], with [membrane]=Nmea(*Area*/*Volume*) being the effective concentration of structured membrane constituents if they were all dissolved in the cell volume.

Regarding (ii), we assume nutrients $A_{nu}$ can diffuse into the cell through a saturating process with rate: $f_{nu}=\mathcal{D}([A_{nu}]_{outside}-[A_{nu}])/(1+[A_{nu}]_{outside}/K_{nu\_outside})$, where $\mathcal{D}$ is an effective diffusion constant (in s$^{-1}$), $[A_{nu}]_{outside}$ is the nutrient concentration in the growth medium outside the cell, and $K_{nu\_outside}$ is a saturation concentration.

## 4.2 Sustained Cellular Growth

The dynamics of the system are now governed by the following equation:

$$dA/dt = Sf + f_{nu} - f_{me} - \mu A$$

where $\mu = f_{me}/[\text{membrane}]$ is the growth rate, and where the term $-\mu A$ represents the dilution factor (concentrations are reduced as cell volume grows). We are now interested in finding if such a system can sustain cellular growth, i.e. if there exists trajectories such that asymptotically $dA/dt = 0$.

Our initial exploration of the ($[\text{membrane}]$, $\mathcal{D}$, $K_{nu\_outside}$, $k_{me}$, $K_{me}$) membrane parameter space suggests the following: neglecting $f_{me}$ saturation (i.e. setting $K_{me}$ arbitrarily large), any numerically accessible parameter set leads to sustained steady-state growth once $[A_{nu}]_{outside}$ is sufficiently large, and nutrient flux is then independent of cytoplasmic state (i.e. $[A_{nu}]_{outside} >> [A_{nu}]$).

Furthermore, there is a broad parameter subspace for which the resulting cytoplasmic state $\{[A_i]\}$ is close to equilibrium. This corresponds to situations where $f_{nu} <<$ metabolic rate (effective transformation rate of $A_{nu}$ in $A_{me}$ by the reaction network). As equilibrium concentrations are an intrinsic property of the system, the cytoplasmic state then only depends on the injected mass flux and is independent of which particular nutrient is injected provided its metabolization rate remains much faster than its injection rate. Figure 5 shows system behavior for such an example parameter set. Cytoplasmic concentrations were found to be close to equilibrium concentrations in the entire $[A_{nu}]_{outside}$ range.

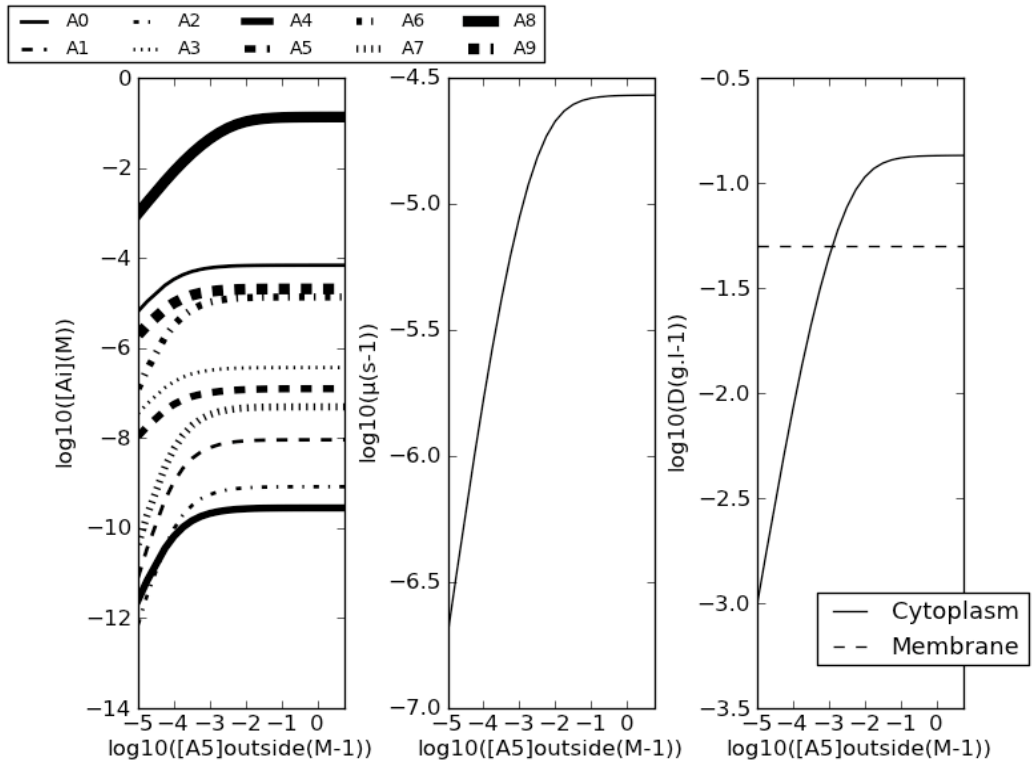

**Figure 5.** Steady-state cytoplasmic concentrations (left), growth rate (center), and cytoplasmic vs. membrane density (right), vs. outside nutrient concentration $[A_5]_{outside}$. Membrane parameters are $\mathcal{D} = 10^{-3}\,\text{s}^{-1}$, $K_{nu\_outside} = 5 \times 10^{-3}\,\text{M}$, $[\text{membrane}] = 5 \times 10^{-2}\,\text{M}$, $k_{me} = 10^{-5}\,\text{s}^{-1}$. Steady-state concentrations are close to equilibrium.

## 5  Discussion and Conclusion

In the range of membrane parameters for which cytoplasmic concentrations are close to equilibrium, our proto-cell is essentially a machine that densifies matter, i.e. that converts a high-energy dilute growth medium (e.g. glucose) into a dense low-energy soup/paste enclosed within a growing membrane. Within a broad range of membrane characteristics, the cytoplasmic soup/paste may remain close to equilibrium, albeit at a larger density than the outside growth medium. Being close to equilibrium grants significant homeostasis: the cytoplasmic state only depends on the injected density flux, and is independent of the particular nutrient provided it is metabolized sufficiently fast.

Thus the growing cell seems to oppose the general principle of diffusion that would tend to equalize concentrations and densities. The apparent contradiction with thermodynamics arises from the fact that we have implicitly assumed that different reactions occur outside vs. inside the cell (glucose is very stable outside the cellular environment). This uncontested observation does not explain how this difference may have arisen in the first place, and this simply points towards the various origin-of-life scenarios [7].

In conclusion, we have shown that complex random chemical reaction networks effectively function as directed transformation systems, producing large amounts of a specific chemical (property of the chemical reaction network) when fed with any nutrient. Assuming this specific chemical can act as a precursor to a structured self-assembled membrane, we have built a proto-cell model capable of sustaining homeostatic cellular growth within a wide range of membrane parameters. In essence, large random conservative chemical reaction networks enclosed within their membrane behave as autopoietic systems, similar to random chemotons but without requiring any explicit informational template [7].

Many aspects of this work require further investigation. In particular, it is presumably the complexity conferred by the maximum size of our random conservative networks that leads to such deterministic and robust qualitative behavior. Characterization of systems below maximum size and as a function of a complexity metric such as R/N would be necessary to verify this point. Formal mathematical demonstrations of computationally observed behavior and/or the development of mean field approaches would also be necessary to extend our conclusions to very large N and R typical of living systems, and that are still beyond computational reach.

## Acknowledgements

We thank Pierre Legrain, Laurent Schwartz and Samuel Bottani for stimulating discussions and advice.